\documentclass[11pt]{article}
\usepackage{vr,amsbsy,amsmath,amsthm,amssymb,times,graphicx,enumerate}
\usepackage[longnamesfirst,sort]{natbib}
\usepackage{animate}
\usepackage{multimedia}
\usepackage{enumerate}
\newcommand{\comments}[1]{}

\newcommand{\Cov}{\mathrm{Cov}}
\newcommand{\Cor}{\mathrm{Cor}}

\newcommand{\sT}{\mathrm{T}}

\begin{document}

\title{A note on marginal correlation based screening}

\author{Run Wang, Somak Dutta and Vivekananda Roy\\
Department of Statistics, Iowa
   State University, USA
}
\date{}
%\corremail{vroy@iastate.edu}

\maketitle

\begin{abstract}
Independence screening methods such as the two
  sample $t$-test and the marginal correlation based ranking are among
  the most widely used techniques for variable selection in ultrahigh
  dimensional data sets. In this short note, simple examples are used
  to demonstrate potential problems with the independence screening
  methods in the presence of correlated predictors. Also, an example
  is considered where all important variables are independent among
  themselves and all but one important variables are independent with
  the unimportant variables. Furthermore, a real data example from a
  genome wide association study is used to illustrate inferior
  performance of marginal correlation screening compared to another
  screening method.
\end{abstract}

{\it Keywords and phrases. correlation, feature selection, screening, sure independence screening, two-sample t test}

\section{Introduction}
\label{sec:int}
Modern scientific research in diverse fields such as engineering,
finance, genetics and neuroimaging involve data sets with hundreds of
thousands of variables. For example, a typical problem in genetics is
to study the association between a phenotype, say, resistance to a
particular disease or yield of plants, and genotype involving millions
of single nucleotide polymorphisms (SNP) markers. Nevertheless, only a
few of these variables are believed to be important. Thus, variable
selection plays a crucial role in modern scientific discoveries.

A variety of methods using different penalizations have been proposed
for variable selection in the linear models, such as the Lasso
\citep{tibs:1996}, SCAD \citep{fan:li:2001}, elastic net
\citep{zou:hast:2005}, adaptive Lasso \citep{zou:2006} and
others. These methods are very useful unless the number of predictors
is much larger than the sample size \citep{wang:2009,
  fan:samw:wu:2009}. In the ultra-high dimensional set up, generally
variable screening is performed to reduce the number of variables
before applying any of the aforementioned variable selection methods
for choosing important variables and estimation of the corresponding coefficients. One widely
used variable screening procedure is based on marginal correlation or
two-sample $t$ test \citep{li:zhon:zhu:2012, fan:lv:2008}. According
to the review article of \citet{saey:inza:larr:2007}, ``the two sample
$t$-test and ANOVA are among the most widely used techniques in
microarray studies'' for feature selection. In particular, one
popular method is to decide which variables should remain in the model
based on ranking of marginal Pearson correlations
\citep{fan:lv:2008}. As mentioned in \cite{fan:lv:2008}, \cite[see
also][]{fan:samw:wu:2009, cho:fryz:2012, clar:clar:2018} there can be several
potential issues with this method, although they are shown to have
sure screening property (that is, with probability tending to one,
important variables survive the screening) under certain conditions.

In this short note, through examples, we illustrate the problems with
marginal correlation based screening in the presence of correlated
predictors. In particular, assuming either autoregressive (order 1) or
equicorrelation covariance structure for normally distributed
predictor variables, it is shown that, with high probability,
important variables do not survive such screening for linear
regression models. We also consider an example where all important
variables are independent among themselves and all but one important
variables are independent with the entire set of unimportant variables. These examples are given in
section~\ref{sec:exam}.  We also consider a real data example from an
agricultural experiment in section~\ref{sec:real} . For this high
dimensional dataset, it is observed that the independence screening
leads to both higher residuals and larger prediction errors than the
high dimensional ordinary least squares projection (HOLP) screening
method of \citet{wang:leng:2016}. Finally, some concluding remarks are
given in section~\ref{sec:concl}.

\section{Examples}
\label{sec:exam}
Suppose the random vector $\mathbf{x} = (x_1,\ldots,x_p)^{\sT}$ has a
multivariate normal distribution with zero mean vector and covariance
matrix $\mathbf{R}.$ Given $\mathbf{x}$ and a random variable
$\epsilon \sim N(0,\sigma^2)$ independent of $\mathbf{x}$, assume that
\begin{equation}\label{eq:regModel}
  y = \beta_0+ \boldsymbol\beta^T \mathbf{x} + \epsilon= \beta_0 + \beta_1x_1 + \cdots + \beta_{p} x_{p} + \epsilon,
\end{equation}
where $\boldsymbol\beta \equiv (\beta_1, \dots, \beta_p)\in \mathbb{R}^{p}$ and all but a few $\beta_i$'s are zero.

In the following examples we shall show that given a covariance matrix
$\mathbf{R},$ the nonzero $\beta_i$'s can take values that would make
marginal correlation between $y$ and important variables smaller in
magnitude than the same between $y$ and each of the unimportant
variables. Indeed, in one example some of the important $x_i$'s become marginally uncorrelated with
the response. We consider here two popular correlation structures. The
first correlation structure is autoregressive where correlation
between $x_i$ and $x_j$ is $\rho^{|i-j|}$ for some $\rho$ between 0
and 1. The second correlation structure is compound symmetric where
the correlation between $x_i$ and $x_j$ is $\rho \in (0,1)$ whenever
$i\ne j.$ 

The general setup of the simulations is as follows. We consider a
particular multivariate Gaussian distribution for the random vector
$\mathbf{x}$ and fix which of the $x_i$'s would have nonzero
coefficients. Then for given value of the correlation parameter, we
choose the values of the nonzero coefficients in such a way that some
of those $x_i$'s become marginally uncorrelated with the response $y$
or have smaller marginal correlation (in absolute value) than that of
the unimportant variables. Then we simulate $n$ independent
realizations from the resulting joint model and compute all the sample
correlations between $y$ and $x_i$, denoted as $ w_i, 1 \le i \le p.$
As mentioned in the introduction, a popular method of screening is to
retain those features with largest absolute marginal correlations,
that is, variables with $\lfloor n/\log(n) \rfloor$ or
$\lfloor n^{1 - \theta} \rfloor$ first largest $|w_i|$ values for some
$0 < \theta <1$ \citep{fan:samw:wu:2009, fan:lv:2008}.  In our
simulation examples, we say a variable does not survive screening if
its absolute marginal correlation is not among the largest $n$ of all.
Based on these $w_i$'s, we observe if a particular subset of important
variables survive screening or not. We repeat this process 100 times
and report the proportion of times the important variables fail to
survive screening.

\paragraph{Example 1}
In this example, we consider the autoregressive correlation design and
show using a simple example that an important variable may fail to
survive screening based on its marginal correlation with the
response. To this end, suppose the $(i,j)$th entry of $\mathbf{R}$ is
$r_{ij} = \rho^{|i-j|}$ where $0 < \rho < 1.$ Note that the largest
eigenvalue of $\mathbf{R}$ is bounded. Next, suppose
$\boldsymbol\beta \in \mathbb{R}^{p}$ be such that $\beta_j = 0$ if
$j \notin \{1,3\}$ and
\[ \begin{pmatrix}
    1 & \rho^2 \\ \rho & \rho
   \end{pmatrix}
   \begin{pmatrix}
    \beta_1 \\ \beta_3
   \end{pmatrix} = 
   \begin{pmatrix}
    0 \\ a
   \end{pmatrix},
\]
where $a \neq 0.$ Then, $\Cov(y,x_1) = 0$ even though $\beta_1 \neq 0$
and $\Cov(y,x_2) = a \neq 0$ even if $\beta_2 = 0.$ For a concrete
example, suppose $\rho = 1/4$ and $a = 3,$ and $\sigma=1.$ Then the
solutions are $\beta_1 = -0.8$ and $\beta_3 = 12.8.$ For a given value
of the sample size $n,$ we set $p = 2n$ and generate data from the
model \eqref{eq:regModel} under the given setup. We repeat this
process 100 times and obtain the proportion of times $x_1$ failed to
survive the screening. In Table~\ref{tab:egAR} we report these
proportions for increasing values of $n.$ Clearly as $n\to \infty,$
variable $x_1$ does not survive screening based on marginal
correlation with non-negligible probability.

In this example, the marginal correlation between $y$ and $x_1$ is
\textit{exactly} zero. In the next example, we consider the
equicorrelation matrix and demonstrate that the important variables
fail to survive screening \textit{even if} the marginal correlations
between $y$ and each of the important variables are bounded away from
zero.

\begin{table}
\begin{center}
\caption{Proportion of times $x_1$ failed to survive screening in Example 1.}
\label{tab:egAR}
 \begin{tabular}{|c|c|c|c|c|} \hline
  $n$  &    50 & 200 & 500 & 1000 \\ \hline
  proportion &  0.56 & 0.45 & 0.44 & 0.62 \\ \hline
 \end{tabular}
 \end{center}
\end{table}

\paragraph{Example 2:} Suppose $\mathbf{R}$ is the equicorrelation
matrix with correlation parameter $\rho.$ That is, the $(i,j)$th
element of $\mathbf{R}$ is $\rho$ if $i\ne j$ and $1$ if $i=j.$ Note
that, the largest eigenvalue of $\mathbf{R}$ is $1 + (p-1)\rho$ which
is $O(n)$ if $p = O(n).$ Without loss, suppose $\beta_i \ne 0, $ for
$i\leq 5$ and $\beta_i = 0$ for $i>5.$ The covariance between $y$ and
$x_i$ is
\[\Cov(y,x_i) = \begin{cases}
                 \beta_i + \rho\sum_{j\ne i} \beta_j & \textrm{ if } \beta_i \ne 0\\
                 \rho\sum_{j=1}^{5}\beta_j& \textrm{ if } \beta_i = 0.
                \end{cases}
\]
Then if we choose $\beta_1,\ldots,\beta_5$ by solving the following system of linear equations
\[\begin{pmatrix}
   1 & \rho & \rho & \rho & \rho \\
   \rho & 1 & \rho & \rho & \rho \\
   \rho & \rho & 1 & \rho & \rho \\
   \rho & \rho & \rho & 1 & \rho \\
   \rho & \rho & \rho & \rho & \rho
  \end{pmatrix}
  \begin{pmatrix}
   \beta_1 \\ \beta_2 \\ \beta_3 \\ \beta_4 \\ \beta_5
  \end{pmatrix} = 
  \begin{pmatrix}
   a \\ a \\ a \\ a \\ b
  \end{pmatrix},
\]
for some $a, b$ with $|a|<|b|$, then we have
\[\Cov(y,x_i) = \begin{cases}
                 a & \textrm{ if } i \leq 4\\
                 b & \textrm{ if } i > 5.
               \end{cases}\] Thus, although $\beta_i \ne 0$
             ($i\leq 4$), the marginal correlations between $y$ and
             each $x_i$ ($i\leq 4$) are all equal but uniformly
             smaller in magnitude than $\Cor(y,x_j), j > 5.$
             Accordingly, based on a sample of size $n$ (where
             $n<<p$), the sample marginal correlation coefficients
             between $y$ and each of $x_1,\ldots,x_4$ will be smaller
             in magnitude than the same between $y$ and each of the
             unimportant variables. Thus, with probability
             tending to one, $x_1,\ldots,x_4$ will fail to survive
             screening with increasing $n$.

             The following simulation study confirms the conclusion.
             We set $a=1, b= 4, p = 2n,$ $\rho = 0.10$ and $\sigma^2 = 1.$ We
             consider $n = 100, 500$ and $1000$. In Table \ref{tab:egEQ} (a) we
             report the proportion of those cases where a particular
             $x_i$ $(i\leq 4)$ failed to survive screening.

             This example is high-dimensional because $p=2n$, but $p$
             increases linearly with $n$. In Table \ref{tab:egEQ} (b)
             we consider the above setup, except with $p = n^2$. We
             see that in this case, the probabilities of not surviving
             the screening increase to one much faster. In
             ultra-high-dimensional cases this problem is further
             exacerbated. However, in ultra-high dimensional problems,
             the largest eigenvalue of the covariance matrix
             $\mathbf{R}$ increases at a larger rate than it does in
             the $p=2n$ case.

\begin{table}
\caption{Proportion of times $x_1,\ldots,x_4$ failed to survive screening in Example 2. Left: $p=2n$, Right: $p = n^2.$}
\label{tab:egEQ}

\begin{center}
  \begin{tabular}{|c|c|c|c|c|} 
  \multicolumn{5}{c}{(a) $p=2n$} \\ \hline
 $n$ & $x_1$ & $x_2$ & $x_3$  & $x_4$ \\ \hline
 100 &   0.68 &  0.61 &  0.59 &  0.54 \\
 500 &   0.83 &  0.86 &  0.90 &  0.92\\
 1000 &  0.96 &  0.96 &  0.96 &  0.97 \\ \hline
 \end{tabular}\hspace{1cm}
 \begin{tabular}{|c|c|c|c|c|} 
 \multicolumn{5}{c}{(b) $p=n^2$} \\ \hline
 $n$ & $x_1$ & $x_2$ & $x_3$  & $x_4$ \\ \hline
25  & 0.95 &  0.98  & 0.96 & 0.98 \\
50  & 0.98 &  0.96  & 0.97 & 0.98 \\
100 & 1.00 &  1.00  & 1.00 & 0.97 \\ \hline 
 \end{tabular}

 \end{center}
\end{table}

\paragraph{Example 3:} Unlike the previous two examples, here we
assume that covariates corresponding to nonzero $\beta_i$'s are
independent. In particular, we assume
\[
  \mathbf{R} = \begin{pmatrix}
    \mathbf{I}_4 & \mathbf{0}^{\top} \\ \mathbf{0} & \mathbf{R}_1
   \end{pmatrix},
 \]
 where $\mathbf{I}_4$ is the $4 \times 4$ identity matrix, $\mathbf{0}$ is the $(p-4) \times 4$
 matrix of zeros, and $\mathbf{R}_1$ is the
 $(p-4) \times (p-4)$ equicorrelation matrix (mentioned in Example 2) with correlation parameter $\rho.$ Also, as in
 Example 2, $\beta_i \ne 0, $ for $i\leq 5$ and $\beta_i = 0$ for
 $i>5.$  Note that, the largest eigenvalue of
 $\mathbf{R}$ is $1 + (p-5)\rho.$ The covariance between $y$ and $x_i$ is
\[\Cov(y,x_i) = \begin{cases}
                 \beta_i & \textrm{ if } i \le 5\\
                 \rho\beta_5& \textrm{ if } i >5.
                \end{cases}
              \]If $\beta_i$'s and $\rho$ are such that for
              $i \le 4$, $|\beta_i| < |\rho\beta_5|$ then the marginal
              correlations between $y$ and $x_i$'s, $i\leq 4$ are all
              smaller in magnitude than $\Cor(y,x_j), j > 5.$ Here,
              all the important covariates are independent among
              themselves and four out of the five true variables are
              independent with all covariates. But the dependence of
              exactly one of the true variables with the unimportant
              variables makes the marginal correlations for rest of
              the important variables uniformly smaller than those for
              the unimportant covariates. (This phenomenon may be
              termed as `nepotism effect' here.) Thus with probability
              tending to one, $x_1,\dots,x_4$ will fail to survive the
              marginal correlation screening with increasing $n$.

              Next, we illustrate the performance of correlation screening using simulation with
              $\rho = 0.2, \beta_i= 1$ for
              $1 \le i \le 4, \beta_5 = 10, \beta_i = 0$ for $i\geq 6$
              and $\sigma^2 = 1$. We consider the same $(n, p)$ values
              as in the Example 2. Based on 100 repetitions
              Table~\ref{tab:egindEQ} provides the proportions of
              those cases where $x_i$ $(i\leq 4)$ failed to survive
              screening. As in Example 2, the probabilities of not
              surviving the screening increase to one faster when
              $p = n^2$ than when $p$ increases linearly with $n$.
    
\begin{table}
\caption{Proportion of times $x_1,\ldots,x_4$ failed to survive screening. Left: $p=2n$, Right: $p = n^2.$}
\label{tab:egindEQ}

\begin{center}
  \begin{tabular}{|c|c|c|c|c|} 
  \multicolumn{5}{c}{(a) $p=2n$} \\ \hline
 $n$ & $x_1$ & $x_2$ & $x_3$  & $x_4$ \\ \hline
 100 &   0.78 &  0.84 &  0.80 &  0.84 \\
 500 &   0.99 &  0.98 &  0.97 &  0.98\\
 1000 &  1.00 &  1.00 &  1.00 &  1.00 \\ \hline
 \end{tabular}\hspace{1cm}
 \begin{tabular}{|c|c|c|c|c|} 
 \multicolumn{5}{c}{(b) $p=n^2$} \\ \hline
 $n$ & $x_1$ & $x_2$ & $x_3$  & $x_4$ \\ \hline
25  & 0.99 &  0.99  & 0.96 & 0.98 \\
50  & 0.99 &  0.98  & 0.99 & 0.98 \\
100 & 1.00 &  1.00  & 1.00 & 1.00 \\ \hline 
 \end{tabular}

 \end{center}
\end{table}

\section{Real data example}
\label{sec:real}
We now study the performance of marginal correlation based screening in a
real dataset.  \citet{cook:mcmu:2012} conducted a genome-wide association
study on starch, protein, and kernel oil content in maize. The
original field trial at Clayton, NC in 2006 consisted of more than
5,000 inbred lines and check varieties primarily coming from a diverse
IL panel consisting of 282 founding lines
\citep{flin:thui:2005}. However, marker information of only
$N = 3,951$ of these varieties are available from the panzea project
(https://www.panzea.org/) which provide information on 546,034 SNPs
after removing duplicates and SNPs with minor allele frequency (MAF)
less than 5\%. We use the phenotype oil content as our response for this
analysis.

\begin{figure*}
  
  \centering\includegraphics[width=5.6in]{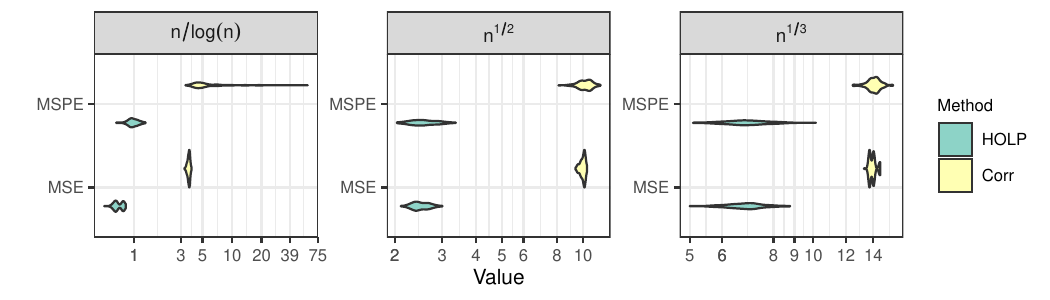}
    \caption{Violin plots of logarithm of MSE and MSPE values.}
    \label{fig:mse}
\end{figure*}

We randomly split the data into a training set of size $n = 3,751$ and
testing set of size 200. Using the training data we select the
variables with large (absolute) marginal correlations. We consider
three different screening model sizes, namely
$\lfloor n/\log(n)\rfloor, \lfloor n^{1/2}\rfloor$ and
$\lfloor n^{1/3}\rfloor$. Then linear models are fit using ordinary
least-squares (OLS) with the selected variables from marginal correlation
screening (Corr) and we compute the mean square errors (MSE) on the training
dataset. Then we calculate the mean square prediction errors (MSPE) on the testing
data based on the OLS fitted  models. We also select variables using HOLP and consider the same
three choices of the model size. Similarly, linear models are fit
using the selected variables from the HOLP. The random splitting and the
whole procedure is repeated 100 times. Figure ~\ref{fig:mse} provides
violin plots of the logarithm of MSE and MSPE values from these
repetitions for both methods. In Figure ~\ref{fig:mse} for the MSPE
plot of independence screening with model size $\lfloor n/\log(n) \rfloor$, we have
dropped an extreme outlier (about $60$) to obtain conspicuous MSE and
MSPE violin plots for the HOLP. From Figure ~\ref{fig:mse} we see that
the independence screening leads to higher
residuals as well as larger MSPEs than the HOLP. Finally, in Table \ref{tab:mse} we report
the median of the MSE and MSPE for each method corresponding to
different selected model sizes. Table \ref{tab:mse} also provides
ratios of these median values, that is,
$\mbox{med}(\mbox{MSE}_{\mbox{HOLP}})/\mbox{med}(\mbox{MSE}_{\mbox{Corr}})$
and
$\mbox{med}(\mbox{MSPE}_{\mbox{HOLP}})/\mbox{med}(\mbox{MSPE}_{\mbox{Corr}})$.
From these ratios we see that inaccuracies in terms of both model fit
and prediction of the marginal correlation screening are exacerbated with
increasing model size.

\begin{table*}
\caption{Median MSE and MSPE for marginal correlation and HOLP screening.}
\label{tab:mse}
\centering

\begin{tabular}{|l|lll|lll|lll|}
\hline
Size   & \multicolumn{3}{c|}{$\lfloor n^{1/3}\rfloor$} & \multicolumn{3}{c|}{$\lfloor n^{1/2}\rfloor$} & \multicolumn{3}{c|}{$\lfloor n/\log(n)\rfloor$} \\\hline
Method & HOLP    & Corr     & Ratio   & HOLP    & Corr     & Ratio   & HOLP    & Corr     & Ratio    \\\hline
MSE    & 6.89    & 14.00   & 0.49    & 2.46    & 10.02   & 0.25    & 0.67    & 3.63    & 0.18     \\
MSPE   & 6.78    & 14.11   & 0.48    & 2.54    & 10.18   & 0.25    & 0.97    & 4.88    & 0.20    \\\hline
\end{tabular}
\end{table*}

\section{Conclusion}
\label{sec:concl}
Here, we study performance of marginal correlation screening for
the linear model with high dimensional data. Correlation
ranking is one of the most widely used techniques for screening out
unimportant features in genetics and other applied sciences. Using
simulation and real data examples, we demonstrate several potential
issues with the independent screening. Since the examples considered
here are fairly simple, we hope that the article can serve the purpose
of providing warning against the use of independence screening such as
the two-sample $t$-test, and the marginal correlation ranking, without
further investigation.

In the presence of nonlinear effects of the covariates on the response,
although not considered in this article, the marginal correlation
screening may miss the true variables \cite[see e.g.][Section 9.1]{clar:clar:2018}. There are several alternatives to the Pearson correlation
screening that have been proposed in the
literature. \citet{fan:lv:2008} proposed the iterated sure independent
screening. Various other correlation measures such as general
correlation \citep{hall:mill:2009}, distance correlation
\citep{li:zhon:zhu:2012}, rank correlation
\citep{li:peng:zhan:zhu:2012}, tilted correlation
\citep{cho:fryz:2012, lin:pang:2014} and quantile partial correlation
\citep{ma:li:tsai:2017} have also been proposed to rank and screen
variables. Thus users may compare the Pearson correlation rankings of
their features with the selected variables from these iterative and
other alternative correlations methods.

\section*{Acknowledgments}
The authors thank the editor for some detailed and careful comments. The authors also thank Ranjan Maitra for some helpful discussions. These comments and discussions have improved the article. Dutta's research was supported in part by the United States Department of Agriculture (USDA) National Institute of Food and Agriculture (NIFA) Hatch project IOW03617. The content presented in this chapter are those of the authors and do not necessarily reflect the views of the NIFA or the USDA.

%\begingroup
%\scriptsize	
% \setstretch{1.0}
\bibliographystyle{WileyNJD-AMS}
\bibliography{ref}
%\endgroup

\end{document}